\documentclass[conference]{IEEEtran}
\IEEEoverridecommandlockouts
\usepackage{cite}
\usepackage{amsmath,amssymb,amsfonts}
\usepackage{algorithmic}
\usepackage{graphicx}
\usepackage{textcomp}
\usepackage{comment}
\usepackage{balance}
\usepackage{xcolor}
\usepackage{booktabs}
\usepackage[absolute]{textpos}

\def\BibTeX{{\rm B\kern-.05em{\sc i\kern-.025em b}\kern-.08em
    T\kern-.1667em\lower.7ex\hbox{E}\kern-.125emX}}
\begin{document}  

\title{Mind the Prompt: Prompting Strategies in Audio Generations for Improving Sound Classification\\ \thanks{*Work carried out during an internship at Bosch Center for Artificial Intelligence, Pittsburgh, USA. 

\footnotesize 
}}

\author{\IEEEauthorblockN{Francesca Ronchini$^{1*}$, Ho-Hsiang Wu$^{2}$, Wei-Cheng Lin$^{2}$, Fabio Antonacci$^{1}$} 

\IEEEauthorblockA{
$^{1}$Dipartimento di Elettronica, Informazione e Bioingegneria (DEIB), Politecnico di Milano, Milano\\
$^{2}$Bosch Center for Artificial Intelligence, Pittsburgh, USA \\
Email: francesca.ronchini@polimi.it, 
ho-hsiang.wu@us.bosch.com,
winston.lin@us.bosch.com, fabio.antonacci@polimi.it
}
}

\maketitle
\begin{abstract}

This paper investigates the design of effective prompt strategies for generating realistic datasets using Text-To-Audio (TTA) models. We also analyze different techniques for efficiently combining these datasets to enhance their utility in sound classification tasks. By evaluating two sound classification datasets with two TTA models, we apply a range of prompt strategies. Our findings reveal that task-specific prompt strategies significantly outperform basic prompt approaches in data generation. Furthermore, merging datasets generated using different TTA models proves to enhance classification performance more effectively than merely increasing the training dataset size. Overall, our results underscore the advantages of these methods as effective data augmentation techniques using synthetic data.

\end{abstract}

\vspace{-0.5em}

\begin{textblock}{14}(1,14.8)
 \footnotesize \noindent © 2025 IEEE.  Personal use of this material is permitted. Permission from IEEE must be obtained for all other uses, in any current or future media, including reprinting/republishing this material for advertising or promotional purposes, creating new collective works, for resale or redistribution to servers or lists, or reuse of any copyrighted component of this work in other works. 
\end{textblock}

\section{Introduction}

Text-to-audio (TTA) models are generative deep learning systems able to generate audio samples from textual information received as inputs, commonly referred to as prompts~\cite{liu2023audioldm, liu2024audioldm, kreuk2022audiogen, majumder2024tango, ghosal2023text, huang2023make, huang2023make2, evans2024fast, evans2024long}. They are encouraging approaches for generating highly realistic synthetic audio datasets, addressing several key challenges of traditional audio collection. Privacy concerns, particularly in applications where real-world audio could infringe on individual rights~\cite{aloufi2020privacy, li2024safeear}, and the limited availability of public domain datasets are significant issues~\cite{mnasri2022anomalous, ma2024foundation}. 
Additionally, specialized audio tasks like anomaly sound detection often lack sufficient datasets~\cite{mnasri2022anomalous, koizumi2020description, pang2021deep}, and the process of labeling audio data is time-consuming, prone to biases and errors, making large-scale, high-quality labeled datasets difficult to produce~\cite{ronchini2021impact, martin2023strong, cornell2024dcase, turpault2019sound}. TTA models are a promising alternative to address these issues by enabling the generation of custom audio samples based on specific requests expressed through natural language, even if fully controlling the generation is challenging~\cite{wu2023audio, lee2024challenge}.

The use of TTA generative models in creating audio datasets is still limited, though previous research has explored this potential across different audio applications~\cite{Ronchini2024, cornell2024generating, kroher2024towards, feng2024can}. In~\cite{kroher2024towards}, the authors propose using TTA models to generate synthetic datasets for music tagging, noting that while synthetic data alone provides limited performance gains, transfer learning and fine-tuning are effective for improving music genre classification. Similar approaches are explored for speech modeling in~\cite{cornell2024generating, feng2024can}, while~\cite{feng2024can} and~\cite{Ronchini2024} focus on using TTA models for Environmental Sound Classification (ESC). All of them demonstrate that TTA-generated data can be used as a valuable data augmentation technique and can effectively replace portions of real data while maintaining state-of-the-art performance~\cite{Ronchini2024, feng2024can}. However, in~\cite{feng2024can} the efficiency of generated datasets is evaluated using pre-trained models, while in~\cite{Ronchini2024} the authors employ older networks on a single dataset, limiting the generalizability of their findings.  Moreover, both studies used basic prompt templates or LLM-guided simple techniques without fully assessing the effectiveness of more advanced prompt strategies. 

Building on prior research and our previous work~\cite{Ronchini2024}, this study makes three significant contributions to the field of sound classification (SC). First, it proposes and analyzes various prompt strategies aimed at efficiently generating datasets. Second, it provides insights into training strategies designed to enhance the performance of datasets generated by Text-To-Audio (TTA) models. Third, it offers a comprehensive analysis of how TTA models can serve as effective alternatives for data augmentation. To achieve these objectives, we select two benchmark SC datasets and generate multiple variations using two advanced TTA models. We introduce three distinct prompt strategies for data generation: a basic template-based approach and two strategies that leverage the Large Language Model (LLM) GPT-4~\cite{achiam2023gpt}. To evaluate the effectiveness of the generated datasets, we train the CNN10 architecture from the PANNs collection~\cite{kong2020panns, wang2019environmental, ding2024acoustic} from scratch, employing combinations of real and synthetic datasets. This comprehensive approach allows us to assess both the quality of the generated data and the effectiveness of different training strategies.

\section{Method}
\label{sec:method}

To effectively address our research goals, we develop three distinct prompt strategies, described in Sec.~\ref{subsec:promptstra}. While initial evaluations demonstrate that these strategies effectively facilitated the creation of relevant prompts, manually crafting individual captions for each audio clip would be impractical and labor-intensive. To streamline this process, we employ a few-shot strategy to engage GPT-4, which enabled us to efficiently generate a complete collection of captions. The captions generation process is detailed in Sec.~\ref{subsec:promptsgens}.

\begin{table}[t!]
\centering
  \caption{Comparison between different prompt strategies used to generate the synthetic dataset to replace the real dataset. The metric reported is accuracy.}
  \footnotesize
  \begin{tabular}{l|c|c|c|c}
  \toprule
  & \multicolumn{2}{c|}{ESC50} & \multicolumn{2}{c}{US8K} \\
  Prompt tech. & Stable Audio & AudioGen & Stable Audio & AudioGen \\
  \midrule
  BSC & 0.34 & 0.30 & 0.39 & 0.42 \\
  STR & 0.40 & 0.26 & \textbf{0.56} & 0.45 \\ 
  EXE & \textbf{0.41} & \textbf{0.31} & 0.51 & \textbf{0.47} \\
  \midrule
  Baseline & \multicolumn{2}{c|}{0.67} & \multicolumn{2}{c}{0.78} \\ 
  \bottomrule
  \end{tabular}
 \label{tab:prompts}
\end{table}

\subsection{Prompt strategies}
\label{subsec:promptstra}

\textbf{Basic prompt strategy}: this strategy uses a single, straightforward instruction to guide the generative model in producing the desired audio. 
It follows the simple and predefined format \textit{``The sound of a $<$sound class$>$"}, where the sound class is replaced with any sound category included in the datasets.

\textbf{Structured prompt strategy}: this approach involves a two-step process. First, we asked GPT-4 to identify key sound attributes for detailed sound descriptions. It proposed five key attributes: pitch, pattern, intensity, acoustic characteristics, and location. In the second step, we prompted GPT-4 to generate natural language sentences that incorporate these attributes. An example of a generated sentence is: \textit{``The quick, high-pitched screech of a chainsaw making short, sharp cuts in softwood."}. This prompt strategy is designed to enrich textual descriptions in a more controlled manner, significantly enhancing audio diversity compared to the basic prompt approach.

\textbf{Exemplar-based prompt strategy}: this strategy leverages human-annotated captions to guide GPT-4 in generating example sentences, which are then used to create the entire dataset. In this paper, we select the Clotho dataset~\cite{drossos2020clotho} as our exemplar dataset, which consists of $6974$ audio clips, each paired with five human-annotated English captions. This approach is motivated by a desire to capture the richness and contextual relevance of human-generated descriptions. By using these annotations as guidance, we aim to ensure that the generated sentences are both high-quality and relevant, enhancing the diversity and accuracy of audio representations in the resulting dataset.

\subsection{Prompts generation}
\label{subsec:promptsgens}

After defining the three prompt strategies, we generate a comprehensive collection of captions for the datasets. In contrast to the Basic prompt strategy, which allows for simple automated generation, the Structured and Exemplar-based strategies require a more nuanced approach. We employ a few-shot methodology by providing GPT-4 with a limited set of randomly selected example captions for each strategy, ensuring that each audio file in the considered datasets receives a unique prompt caption. For the Structured strategy, these examples consist of sentences generated by GPT-4, while for the Exemplar-based strategy, we select examples directly from the Clotho dataset. To create the entire collection of captions, we instruct GPT-4 to use these examples as a foundation for generating new captions for all audio files. This process includes providing the model with detailed instructions to ensure it generates varied and original captions that emphasize creativity and diversity, while also being well-structured and relevant to their respective sound classes.

\begin{table}[t!]
\centering
  \caption{Comparison between different prompt strategies when generated datasets are used as data augmentation technique.
  The metric reported is accuracy.}
  \footnotesize
  \begin{tabular}{l|c|c|c|c}
  \toprule
  & \multicolumn{2}{c|}{ESC50} & \multicolumn{2}{c}{US8K} \\
  Prompt tech. & Stable Audio & AudioGen & Stable Audio & AudioGen \\
  \midrule
  BSC w/ ORG & 0.70 & 0.67 & 0.77 & 0.78  \\
  STR w/ ORG & \textbf{0.72} & 0.67 & \textbf{0.79} & \textbf{0.78} \\
  EXE w/ ORG & 0.69 & \textbf{0.68} & \textbf{0.79} & \textbf{0.78} \\
  \midrule
  Baseline & \multicolumn{2}{c|}{0.67} & \multicolumn{2}{c}{0.78} \\ 
  \bottomrule
  \end{tabular}
 \label{tab:da}
\end{table}

\section{Experimental design}
\label{sec:expdesign}

\subsection{Audio Generative Models}

We use two pre-trained TTA models for data generation. \textbf{AudioGen (AG)}~\cite{kreuk2022audiogen} is an auto-regressive model that encodes raw audio into a discrete representation and generates audio using a transformer conditioned on text. \textbf{Stable Audio Open (SA)}~\cite{evans2024stable} is a latent diffusion model that produces variable-length stereo audio from text prompts. It consists of an autoencoder for waveform compression, a T5-based text embedding, and a transformer-based diffusion model (DiT) for audio generation. 
AudioGen is selected for its strong performance in previous studies~\cite{Ronchini2024, feng2024can}, while Stable Audio Open for its promising results~\cite{evans2024stable}. For a fair comparison, the output of Stable Audio Open is resampled to $16~\mathrm{kHz}$ and converted to mono.

\subsection{Datasets}
We select two well-known SC datasets used as benchmarks in the literature~\cite{bansal2022environmental, nogueira2022sound, bhattacharya2021deep}. \textbf{Environmental Sound Classification (ESC50)}~\cite{piczak2015dataset} contains $2000$ environmental audio recording of $5\mathrm{s}$ length divided in $50$ sound categories, each containing $40$ audio samples. \textbf{UrbanSound8k (US8K)}~\cite{salamon2014dataset} is composed of $8732$ labeled sounds of $4\mathrm{s}$ maximum duration of urban sounds divided into $10$ classes. 

Each copy of the datasets is generated following the original dataset distributions.

\subsection{Model Architecture and Evaluation}
We use CNN10~\cite{kong2020panns} to evaluate the study performances, as CNNs are widely used for audio tagging and sound classification~\cite{kong2020panns, mohmmad2024exploring, chan2020comprehensive, nogueira2022sound}. In each experiment, the network is trained from scratch for up to $200$ epochs, with early stopping applied based on validation loss, using patience of 10 epochs. The SC model performances are evaluated using accuracy as the main metric since we are considering balanced datasets. For the ESC50 dataset, $5$-fold cross-validation is applied, while for US8K, $10$-fold cross-validation is used. The evaluation follows the same cross-validation distribution as the original datasets~\cite{piczak2015dataset, salamon2014dataset}. We consider the CNN10 trained with original datasets as the baseline.

\section{Experiments and results}
\label{sec:exps}
This section presents a comprehensive overview of the experiments conducted in this paper, together with their corresponding results. In the tables, the Basic strategy is referred to as BSC, the Structured strategy as STR, and the Exemplar-based strategy as EXE. ORG stands for original dataset. We informally computed the confusion matrices and for all experiments. While we do not report them in this paper due to the preliminary nature of the results, which require further in-depth analysis, we offer some general observations derived from these matrices to provide initial insights into the significance of the findings.

\subsection{How effective are different prompt strategies in enhancing performance?}
\label{subsec:prompts}

\begin{figure}[t!]
\centering
\begin{minipage}[b]{1.0\columnwidth}  
  \centering
  \includegraphics[width=0.9\linewidth]{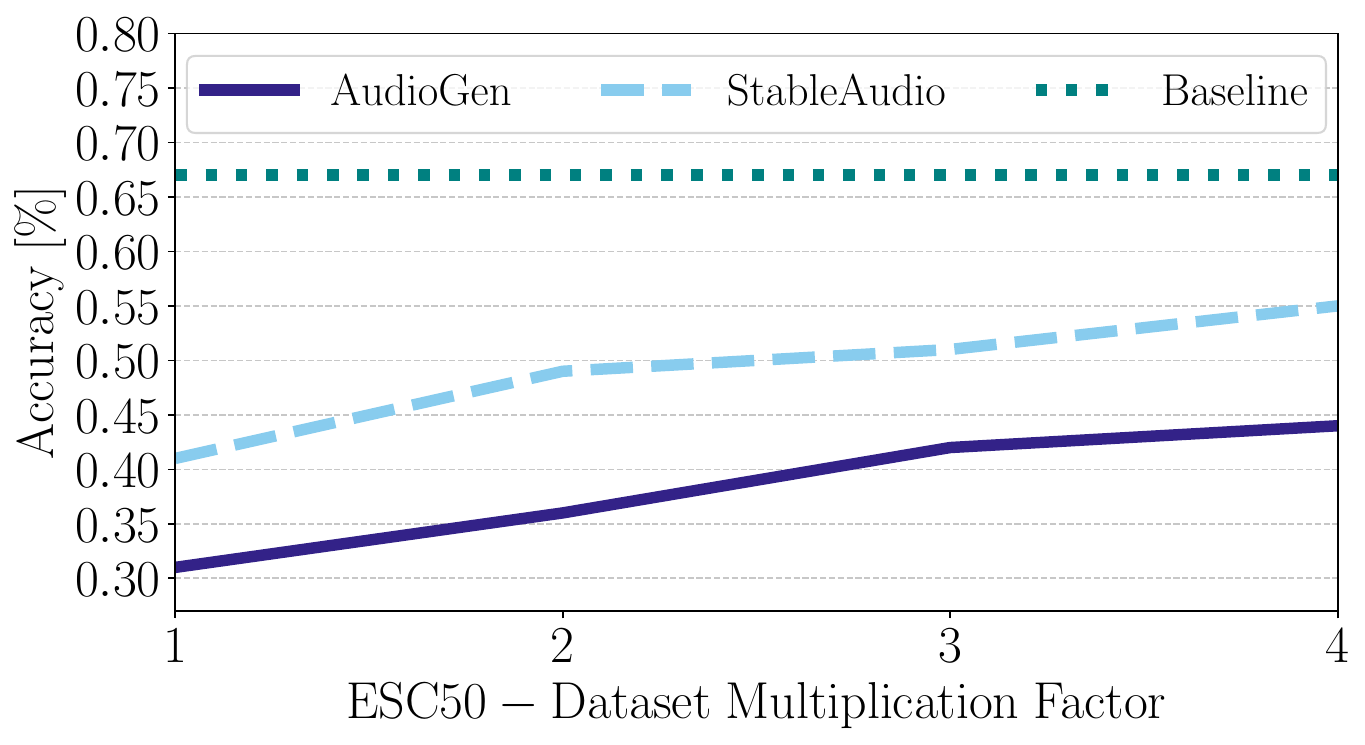}
  \centerline{(a)}
  \medskip
\end{minipage}
\begin{minipage}[b]{1.0\columnwidth}  
  \centering
  \includegraphics[width=0.9\linewidth]{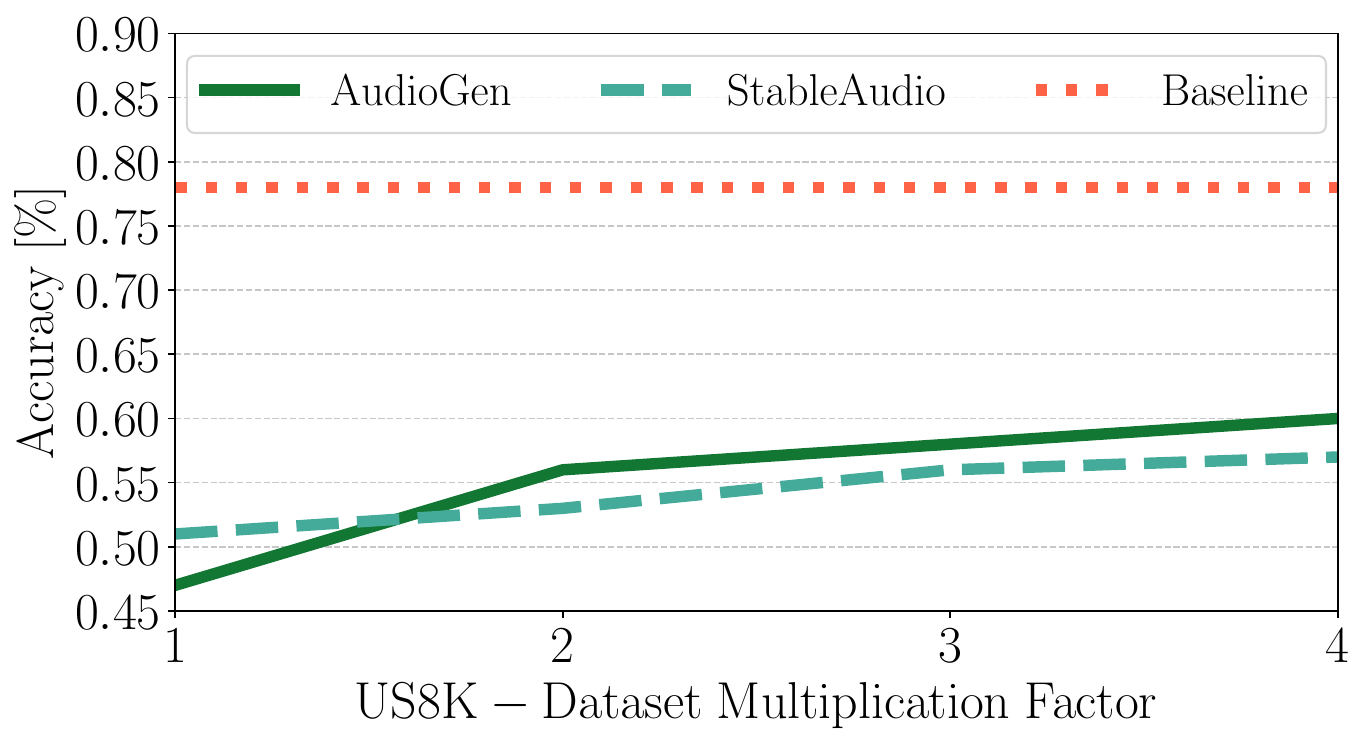}
 \centerline{(b)}
  \medskip 
\end{minipage}
\caption{Accuracy of CNN10 when trained with ESC50 (a) and US8K (b) TTA-generated datasets.}
\label{fig:onlyfake}
\end{figure}

This experiment investigates the impact of different prompt strategies on performance when training CNN10 exclusively with synthetic data. As shown in Table~\ref{tab:prompts}, task-specific prompt strategies lead to significant improvements over the Basic strategy. However, while detailed textual prompts enhance the quality and diversity of TTA output, achieving complete control over the generation process remains a challenge. This limitation affects the potential for state-of-the-art performance in dataset generation using TTA models. Additionally, the observed results may be influenced by data distribution mismatch between the training and testing datasets. Further investigation is needed to confirm this hypothesis.
Informal results report that similar sounds, whether from human activities (e.g., sneezing and coughing), perceptual overlaps (e.g., helicopters and airplanes), or shared environments (e.g., sheep and cows), are often confused. This may be due to co-occurrence in TTA training clips or difficulty in distinguishing similar acoustic features. Moreover, while real-world datasets are properly curated, TTA generative models often mix related sounds (e.g., keyboard clicks with mouse clicks), which sometime appear in the same sound clip when listening to some samples. 

\begin{figure}[t!]
\centering
\begin{minipage}[b]{1.0\columnwidth}  
  \centering
  \includegraphics[width=0.9\linewidth]{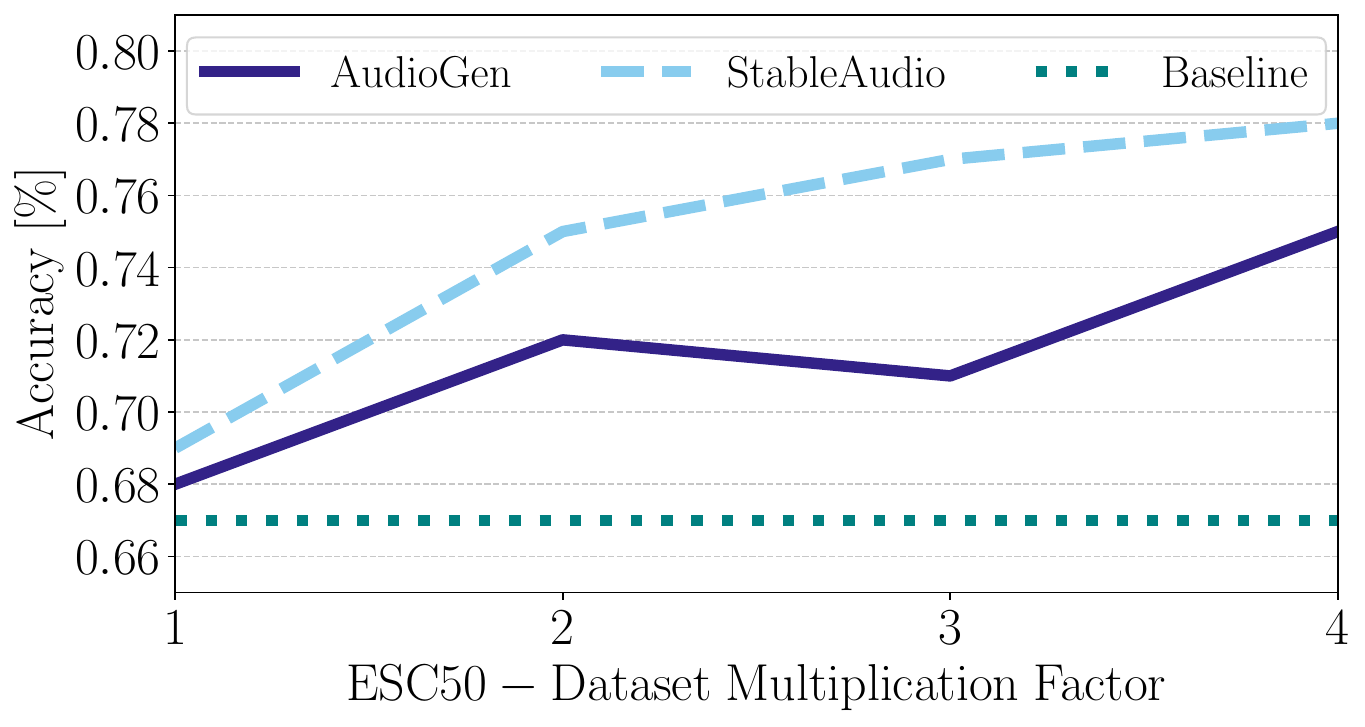}
  \centerline{(a)}
  \medskip
\end{minipage}
\begin{minipage}[b]{1.0\columnwidth}  
  \centering
  \includegraphics[width=0.9\linewidth]{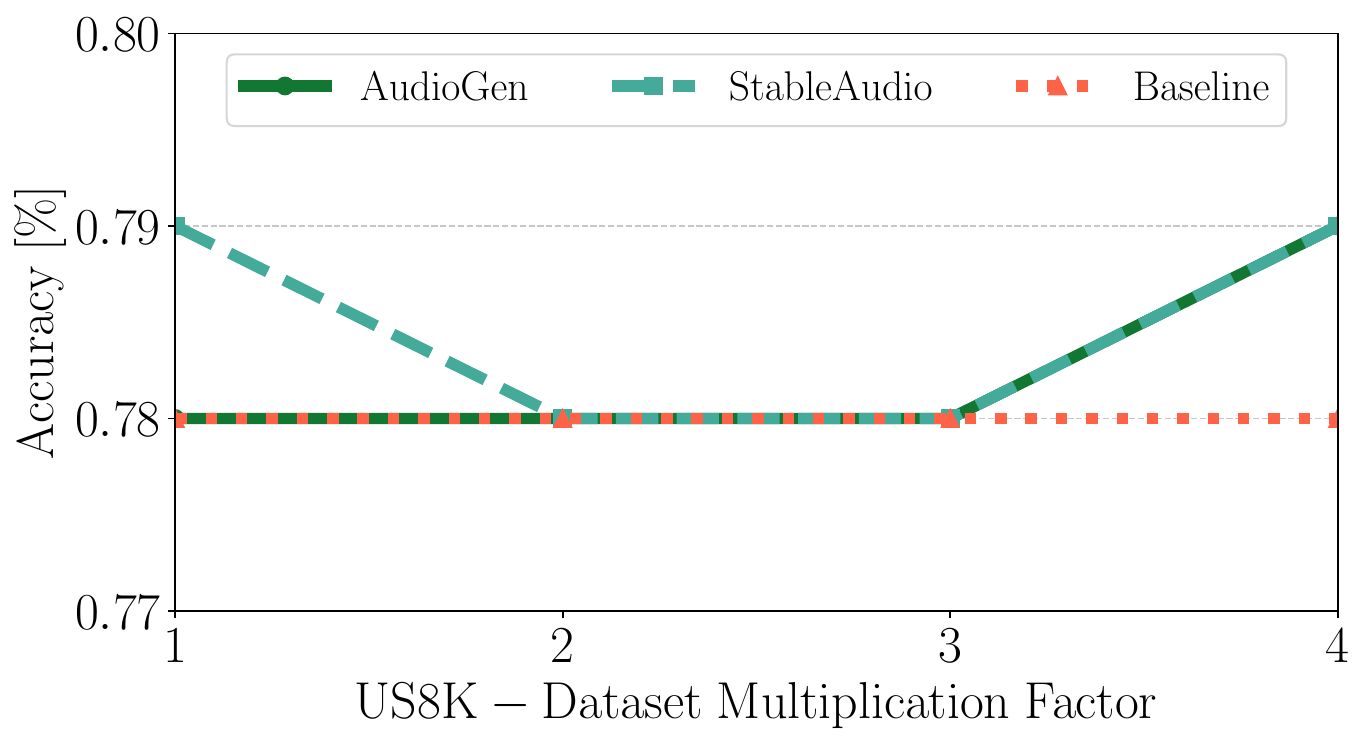}
 \centerline{(b)}
  \medskip
\end{minipage}
\caption{Accuracy of CNN10 when ESC50 (a) and US8K (b) TTA-generated datasets are used as a data augmentation technique.}
\label{fig:da}
\end{figure}

\subsection{Are different prompt strategies also effective in enhancing performance when generated data are used as augmentation?}
\label{subsec:da}

Previous studies have shown that while TTA-generated data may not yet achieve baseline performance, they can serve as an effective data augmentation technique \cite{Ronchini2024, feng2024can}. Building on these findings, we replicate the previous experiment incorporating the original dataset during training. The results in Table~\ref{tab:da} confirm that using enriched and detailed prompt strategies, such as the Structured or Exemplar-based, increases the diversity of the TTA-generated data. However, this improvement does not extend to the US8K dataset when generated using AudioGen. The gap in performances could be explained by the dataset's distribution differences: ESC50 contains fewer samples across a larger number of categories, while US8K has more samples distributed among fewer classes. Moreover, when listening to the original recordings, ESC50 samples are clear, single sound source files (e.g. a clear sound of a dog barking), while most of the US8K audio clips have background noise and, often, multiple sound sources are overlapped 
(e.g. a dog barking and people talking at the same time). This complexity makes describing the desired audio more challenging compared to single-source files and increases the difficulty of generating the audio accurately.

\subsection{Does increasing the number of training files lead to improved performance?}

For this experiment, we only consider the EXE strategy, as it has proven to be the most promising prompt strategy among those designed for study. We increase the number of generated files by $2$x, $3$x, and $4$x to train the CNN10 network. Fig.~\ref{fig:onlyfake}(a) and Fig.~\ref{fig:onlyfake}(b) present the results when the network is trained solely on TTA-generated data, while Fig.~\ref{fig:da}(a) and Fig.~\ref{fig:da}(b) illustrate the same results when using TTA-generated datasets as a data augmentation approach. For comparison, the figures also show the accuracy for a single copy of the dataset ($1$x). Both scenarios report a similar trend, even if with a reduced effect in the data augmentation case. Increasing the number of files at training narrows the performance gap between synthetic and real data, although baseline performances are not achieved. Additionally, we observe different behaviors between the generative models across the datasets.
This may be due to the generative models being trained on different datasets~\cite{evans2024stable, kreuk2022audiogen}, which likely affects the audio quality and generalization capabilities of the SC models, contributing to an increasing distribution mismatch between the training and testing data. Our informal observations also indicate persistent class confusion; despite an increase in the number of files, certain classes, such as sheep and cows, continue to be misidentified, highlighting ongoing confusion between them.

\subsection{Can the performance be improved by mixing various prompt strategies?}
\label{subsec:promptsmix}

\begin{table}[t!]
\centering
  \caption{Accuracy of CNN10 when trained with merged datasets generated from different prompt techniques.}
  \footnotesize
  \begin{tabular}{l|c|c|c|c}
  \toprule
  & \multicolumn{2}{c|}{ESC50} & \multicolumn{2}{c}{US8K} \\
  Prompt technique(s) & SA & AG & SA & AG \\
  \midrule
  BSC, EXE & 0.48 & 0.40 & 0.50 & 0.51  \\
  BSC, STR & 0.46 & 0.38 & 0.53 & 0.56 \\
  EXE, STR & 0.48 & 0.36 & \textbf{0.57} & 0.55 \\
  BSC, EXE, STR & \textbf{0.55} & \textbf{0.42} & 0.55 & \textbf{0.60} \\
  BSC, EXE, STR w/ ORG & 0.75 & \textbf{0.76} & 0.78 &  \textbf{0.79} \\
  \midrule
  Baseline & \multicolumn{2}{c|}{0.67} & \multicolumn{2}{c}{0.78} \\ 
  \bottomrule
  \end{tabular}
 \label{tab:mixprompts}
\end{table}

While the previous experiments provide valuable insights, they do not completely clarify the performance differences. A crucial question remains: are the observed performance improvements primarily driven by the prompt strategies, or do they arise from the increased number of training files? To investigate this further, we design an additional experiment that combines various prompt strategies, training the models with a mix of these approaches. The results, as presented in Table~\ref{tab:mixprompts}, reveal an interesting finding: employing two different prompt strategies to generate data enhances the performance of the SC models more effectively than merely increasing the data volume with a single prompt, showing that data diversity is key to model improvement. This is emphasized comparing these results to Fig.~\ref{fig:onlyfake} and Fig.~\ref{fig:da}, underscoring the importance of varied prompts in achieving diverse training outcomes over a large, uniform dataset.

\subsection{Does merging outputs from different generative models enhance overall results?}
\label{subsec:modelsmix}

Building on our findings, we hypothesize that the diversity in training methods and dataset distributions among TTA models may yield unique data representations, influencing sound generation and emphasizing different aspects of the data. To gain further insights, we train CNN10 on merged synthetic datasets generated using the same prompt strategy across various TTA models. The results in Table~\ref{tab:mixmodels} show that combining datasets created by different TTA models leads to better performance than just doubling the data from a single TTA model. This conclusion is further supported when comparing the results with the ones reported in Fig.~\ref{fig:da}. 
The findings support our hypothesis that different TTA models can capture diverse data characteristics and a mixed dataset leverages the strengths of each of them.

\begin{table}[t!]
\centering
  \caption{Accuracy of CNN10 when trained on merged datasets generated using the same prompt technique across different TTA models.}
  \footnotesize
  \begin{tabular}{l|c|c}
  \toprule
  Prompt technique & ESC50 (SA + AG) & US8K (SA + AG) \\
  \midrule
  EXE & \textbf{0.52} & 0.60 \\
  STR & 0.49 & \textbf{0.61} \\
  \midrule
  EXE w/ ORG & 0.75 & 0.79\\
  STR w/ ORG & 0.75 & \textbf{0.80} \\
  \midrule
  Baseline &  0.67 & 0.78 \\ 
  \bottomrule
  \end{tabular}
 \label{tab:mixmodels}
\end{table}

\section{Conclusions and future works}
\label{sec:con}

This paper proposes and analyzes different prompt strategies for generating captions aimed at efficiently creating datasets as substitutes or data augmentation strategies for original datasets in sound classification applications. The study further investigates different dataset combinations to optimize the use of TTA technology for audio sample generation. The results provide valuable insights into effectively prompting TTA models for synthetic data generation. Future work will focus on fine-tuning to enhance TTA model capabilities and explore domain adaptation methods to improve the generalization of SC models when trained with generated datasets. Also, using an LLM for caption generation may lead to unintentionally biased or repetitive outputs. Additionally, with the Structured strategy, we cannot guarantee that GPT-4 consistently includes all five attributes in every generated sentence. Future research will explore strategies to mitigate these limitations. 

\newpage
\bibliographystyle{IEEEtran}
\bibliography{biblio}

\end{document}